\documentclass{appolb}
\usepackage{graphicx}
\usepackage{braket}
\usepackage{amsmath}
\usepackage{tikz}
\newcommand{\kaon}{\mathrm{K}^0}
\newcommand{\akaon}{\bar{\mathrm{K}}^0}

\newcommand{\Ks}{\mathrm{K_S}}
\newcommand{\Kl}{\mathrm{K_L}}

\newcommand{\Kp}{\mathrm{K_{+}}}
\newcommand{\Km}{\mathrm{K_{-}}}
\newcommand{\Ts}{$\mathcal{T}$}
\newcommand{\CPs}{$\mathcal{CP}$}
\newcommand{\CPTs}{$\mathcal{CPT}$}
\begin{document}
%
\title{
Study of the $\Ks\Kl\to\pi\ell\nu 3\pi^0$ process\\for time reversal symmetry test at KLOE-2
}
\author{Aleksander Gajos\\
on behalf of the KLOE-2 collaboration
\address{The Marian Smoluchowski Institute of Physics, Jagiellonian University,\\
\L{}ojasiewicza 11, 30-348, Krak\'ow, Poland\\
aleksander.gajos@uj.edu.pl
}
\\
}
\maketitle
\begin{abstract}
This work presents prospects for conducting a novel direct test of time-reversal symmetry at the KLOE-2 experiment. Quantum entanglement of neutral K meson pairs uniquely available at KLOE-2 allows to probe directly the time-reversal symmetry ($\mathcal{T}$) independently of $\mathcal{CP}$ violation. This is achieved by a comparison of probabilities for a transition between flavour and $\mathcal{CP}$-definite states and its inverse obtained through exchange of initial and final states. As such a test requires the reconstruction of the $K_L\to 3\pi^0$ decay accompanied by $K_S\to \pi^{\pm}\ell^{\mp}\nu$ with good timing information, a new reconstruction method for this process is also presented which is capable of reconstructing the $K_L\to 3\pi^0$ decay with decay time resolution of $\mathcal{O}(1\tau_S)$.
\end{abstract}
  
\section{Introduction}
Well known for \CPs-violating phenomena, neutral kaons may also be used to study directly the time-reversal symmetry although special care is necessary to prepare a \Ts~symmetry test which should be independent of \CPs-violation effects. Such a test is possible with entangled neutral kaon pairs uniquely available at the DA$\Phi$NE $\phi$-factory \cite{Bernabeu:2012nu}. Kaon transitions between flavour-definite and \CPs-definite states constitute processes for whom an exchange of initial and final state only corresponds to the time-reversal operation and not \CPs~nor \CPTs~conjugation. This allows for a direct test by comparison of amplitudes for a transition and its inverse independently of \CPs~and \CPTs. A similar principle was recently used by the BaBar experiment to observe \Ts-violation in the neutral B meson system \cite{Bernabeu:2012ab,Lees:2012uka}. In turn, KLOE-2 is capable of investigating time-reversal violation with neutral kaons.

\section{The direct \Ts~symmetry test}
\begin{figure}[htb]
  \centerline{%
    \includegraphics[width=8cm]{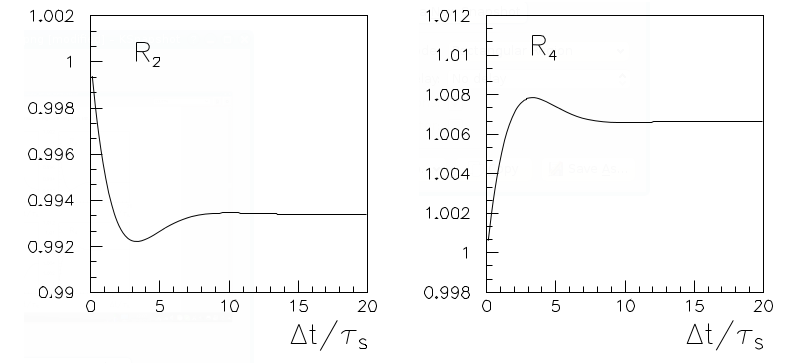}
  }
  \caption{Expected behaviour of the transition probability ratios $R_2$ and $R_4$ as a function of proper decay times difference $\Delta t$ as simulated for 10$fb^{-1}$ of KLOE-2 data. Figure adapted from \cite{Bernabeu:2012nu}.}
  \label{Fig:ratios}
\end{figure}

The entangled states of a pair of neutral K mesons produced in the $\phi$ meson decay may be expressed in any suitable basis of orthogonal states such as flavour-definite states $\{\kaon,\akaon\}$ or \CPs-definite states $\{\Kp,\Km\}$:
\begin{equation}
    \Ket{\phi} \to \frac{1}{\sqrt{2}} \left( \ket{\kaon}\ket{\akaon}-\ket{\akaon}\ket{\kaon} \right) = \frac{1}{\sqrt{2}} \left( \Ket{\Kp}\Ket{\Km}-\Ket{\Km}\Ket{\Kp} \right).
\end{equation}
Kaons can be identified in these bases through final state observation at the moment of their decay. If the $\Delta S = \Delta Q$ rule is assumed\footnote{The $\Delta S=\Delta Q$ rule is well tested in semileptonic kaon decays\cite{Beringer:1900zz}}, the semileptonic decays with a positively and negatively charged leptons (later denoted as $\ell^+$, $\ell^-$) unambiguously tag the decaying state as $\kaon$ and $\akaon$. Meanwhile, hadronic decay modes with two and three pions ($3\pi^0$)\footnote{Only $3\pi^0$ is a pure CP=-1 state.} are only possible for \CPs~eigenstates $\Kp$ (CP=1) and $\Km$ (CP=-1), respectively.
Observation of a transition between \CPs~and flavour-definite states also requires identification of kaon state at a point before its decay. This is uniquely possible with entangled neutral kaon pairs, as recognition of the state of the first decaying kaon guarantees its still-living partner to be in the orthogonal state at the moment of the first decay. Therefore it is possible to obtain the transitions listed in table~\ref{Tab:trs} along with their time inverses. It is worth stressing that these transitions are connected with their \Ts-inverses only by time-reversal conjugation and not by \CPs~nor \CPTs~transformations.
\begin{table}[h]
  \centering
  \begin{tabular}{r|cl|cl}
    &  Transition &  & $\mathcal{T}$-conjugate &  \\\hline
    1 &  $\kaon \to \Kp$ & $(\ell^-,\pi\pi)$ & $\Kp \to \kaon$ & $(3\pi^0, \ell^+)$\\
    2 &  $\kaon \to \Km$ & $(\ell^-,  3\pi^0)$  &$\Km \to \kaon$ & $(\pi\pi,\ell^+)$\\
    3 &  $\akaon \to \Kp$ & $(\ell^+,\pi\pi)$ & $\Kp \to \akaon$ & $(3\pi^0, \ell^-)$\\
    4 &  $\akaon \to \Km$ & $(\ell^+,3\pi^0)$ & $\Km \to \akaon$ & $(\pi\pi, \ell^-)$
  \end{tabular}
  \caption{Transitions between flavour and CP-definite states of neutral kaons. For each transition a time-ordered pair of final states indicating the decays of respective states is provided in parentheses.}
  \label{Tab:trs}
\end{table}
For each of the transitions from table~\ref{Tab:trs} a measurement of the ratio of time-dependent probabilities of a transition and its inverse constitutes a test of \Ts~symmetry. At KLOE-2 \cite{AmelinoCamelia:2010me} statistically significant tests are expected for transitions 2 and 4. The theoretical ratios $R_2$ and $R_4$ can be experimentally obtained from measurable ratios of double decay rates to which they are proportional up to a constant: 
\begin{align}
  {R_2(\Delta t)} & = {\frac{P[\kaon(0) \to \Km(\Delta t)]}{P[\Km(0) \to \kaon(\Delta t)]} } \; \sim \; \frac{\mathrm{I}(\ell^-,3\pi^0;\Delta t)}{\mathrm{I}(\pi\pi,\ell^+;\Delta t)}, \\
  {R_4(\Delta t)} & = {\frac{P[\akaon(0) \to \Km(\Delta t)]}{P[\Km(0) \to \akaon(\Delta t)]} } \; \sim \; \frac{\mathrm{I}(\ell^+,3\pi^0;\Delta t)}{\mathrm{I}(\pi\pi,\ell^-;\Delta t)},
\end{align}
where $\Delta t$ is the difference of proper decay times of the two kaons. Any discrepancy of the $R_2$ and $R_4$ ratios from unity would be a direct signal of \Ts~symmetry violation. At KLOE-2 the asymptotic behaviour of these ratios can be measured (see Fig.~\ref{Fig:ratios}) in order to extract the \Ts-violating $Re(\epsilon)$ parameter as the theoretical prediction for large time differences is $R_2(\Delta t) \stackrel{\Delta t \gg \tau_s}{\longrightarrow} 1-4Re(\epsilon)$ and $R_4(\Delta t) \stackrel{\Delta t \gg \tau_s}{\longrightarrow} 1+4Re(\epsilon)$.
\section{Experimental realization at KLOE-2 and DA$\Phi$NE}
\label{Sec:experiment}
The DA$\Phi$NE $\phi$-factory is an electron-positron collider operating at the energy of the $\phi$ resonance peak ($\sqrt{s}\approx 1020$MeV) and predominantly producing $\phi$ mesons with small momentum ($\beta_{\phi}\approx 0.015$) whose decays provide pairs of charged or neutral kaons with branching fractions of about 49\% and 34\% respectively. Kaon decays are recorded by the KLOE detector consisting of a cylindrical drift chamber (DC) surrounded by a sampling electromagnetic calorimeter (EMC). In the recent upgrade to KLOE-2, the region close to interaction point was filled with a novel Cylindrical triple-GEM inner tracker (IT) to improve vertexing \cite{Balla:2013gua}.

As shown in the previous section, a direct test of \Ts~symmetry at KLOE-2 requires ability to reconstruct two types of events: $\Ks\Kl\to \ell^{\pm}\pi^{\mp}\nu\:3\pi^0$ and $\Ks\Kl\to \pi\pi\:\ell^{\pm}\pi^{\mp}\nu$. For construction of time-depedent decay distributions, kaon proper decay times should be determined with resolution of the order of 1 $\tau_S$. In case of $\pi^+\pi^-$ (chosen as the $\pi\pi$ state) and semileptonic final states, charged particle tracks provide good vertexing (and thus timing) information. The $\Kl\to 3\pi^0\to 6\gamma$ decay, however, is a challenging reconstruction task as only neutral particles are involved and the only recorded information on this process are the $\gamma$ hits in the EMC. For this decay a new reconstruction method was prepared for KLOE-2.
\section{$\Kl \to 3\pi^0$ decay reconstruction}
\label{Sec:reconstruction}
\begin{figure}[h]
  \centering
  \definecolor{gray1}{gray}{0.8}
  \definecolor{gray2}{gray}{0.7}
  \definecolor{gray3}{gray}{0.5}
  \definecolor{gray4}{gray}{0.3}
  \definecolor{gray5}{gray}{0.1}
\vspace{-0.5cm}
  \begin{tikzpicture}[scale=0.35]
    \coordinate (clu1) at (-4.58,2);
    \coordinate (clu2) at (1,4.90);
    \coordinate (clu3) at (4,3);
    \coordinate (clu4) at (4.90,-1);
    \coordinate (clu5) at (-2,4.58);
    \coordinate (clu6) at (4.90, 1.0);
    \coordinate (vertex) at (1.0,2.0);
    \node [] (upperphantom) at (-6,5) {};
    
    \draw[gray2, line width=3pt] (0,0) circle (5);
    \draw[gray4,fill] (clu1) circle (0.2);
    \draw[gray4,fill] (clu2) circle (0.2);
    \draw[gray4,fill] (clu3) circle (0.2);
    \draw[gray4,fill] (clu4) circle (0.2);
    \draw[gray4,fill] (clu5) circle (0.2);
    \draw[gray4,fill] (clu6) circle (0.2);
    
    \node[black,anchor=south] at (clu1) {\scriptsize$(X_1,Y_1,Z_1,T_1)$};
    \draw[gray4,rotate=85, dashed, thick] (clu1)+(-5.58,0) arc (180:320:5.58);
    
    \draw[thick, gray3,->] (0.63, 0) -- (-4.3,1.9) node[midway, above] {$\gamma$} node[midway, below, black] {\footnotesize $c(T_1-t)$};
  \end{tikzpicture} \hspace{1cm}
  \begin{tikzpicture}[scale=0.35]
    \coordinate (clu1) at (-4.58,2);
    \coordinate (clu2) at (1,4.90);
    \coordinate (clu3) at (4,3);
    \coordinate (clu4) at (4.90,-1);
    \coordinate (clu5) at (-2,4.58);
    \coordinate (clu6) at (4.90, 1.0);
    \coordinate (vertex) at (1.0,2.0);
    \node [] (upperphantom) at (-6,5) {};
    
    \draw[gray2, line width=3pt] (0,0) circle (5);
    \draw[gray4,fill] (clu1) circle (0.2);
    \draw[gray4,fill] (clu2) circle (0.2);
    \draw[gray4,fill] (clu3) circle (0.2);
    \draw[gray4,fill] (clu4) circle (0.2);
    \draw[gray4,fill] (clu5) circle (0.2);
    \draw[gray4,fill] (clu6) circle (0.2);
    
    \draw[thick, gray4,rotate=85, dashed] (clu1)+(-5.58,0) arc (180:320:5.58);
    
    \draw[thick, gray4,rotate=-10,dashed] (clu2)+(-2.90,0) arc (180:360:2.90);
    \draw[thick, gray4,rotate=-55,dashed] (clu3)+(-3.16,0) arc (180:360:3.16);
    \draw[thick, gray4,rotate=-90, dashed] (clu4)+(-4.92,0) arc (180:330:4.92);
    \draw[thick, gray4,rotate=20,dashed] (clu5)+(-3.96,0) arc (180:360:3.96);
    \draw[thick, gray4,rotate=-60, dashed] (clu6)+(-4.03,0) arc (180:330:4.03);
    \draw[thick, black, fill] (vertex) circle (0.1) node[left]{\footnotesize $(x,y,z,t)$\phantom{a}};
  \end{tikzpicture}
  \caption{Scheme of the $\Kl\to 3\pi^0\to 6 \gamma$ decay vertex reconstruction in cross-section view of the KLOE EMC (grey ring). 
  }
  \label{Fig:scheme}
\end{figure}
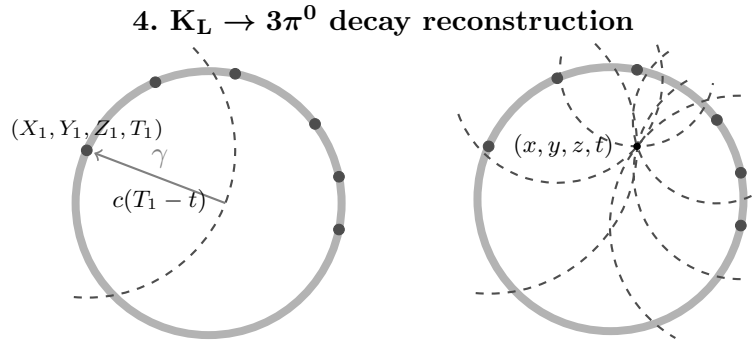
The new reconstruction procedure uses only information on up to 6 $\gamma$ hits in the KLOE-2 EMC in order to reconstruct both spatial location and time of the $\Kl\to 3\pi^0\to 6 \gamma$ decay. For each of the photons, EMC provides information on the hit point and time (Fig.~\ref{Fig:scheme}, left). Therefore a set of possible origin points of the incident $\gamma$ is a sphere centered at the EMC hit position with a radius dependent on the time of the $\Kl$ decay \textit{t}. Such spheres for each available EMC $\gamma$ hit constitute a system of equations:
\begin{equation}
  (T_i-t)^2c^2 = (X_i-x)^2+(Y_i-y)^2+(Z_i-z)^2 \quad  i=1,\ldots,6.
\end{equation}
As the $\Kl$ decay vertex is the common origin of all photons, it can be found as an intersection of all spheres defined above by solving the system of equations for \textit{x,y,z} and \textit{t} (Fig.~\ref{Fig:scheme}, right). Although only 4 $\gamma$ hits are necessary to solve the system, recording all 6 photons allows to improve the decay vertex resolution by numerical best satisfaction of the overdetermined system. 

\begin{figure}[htb]  \centerline{%
    \includegraphics[width=7.0cm]{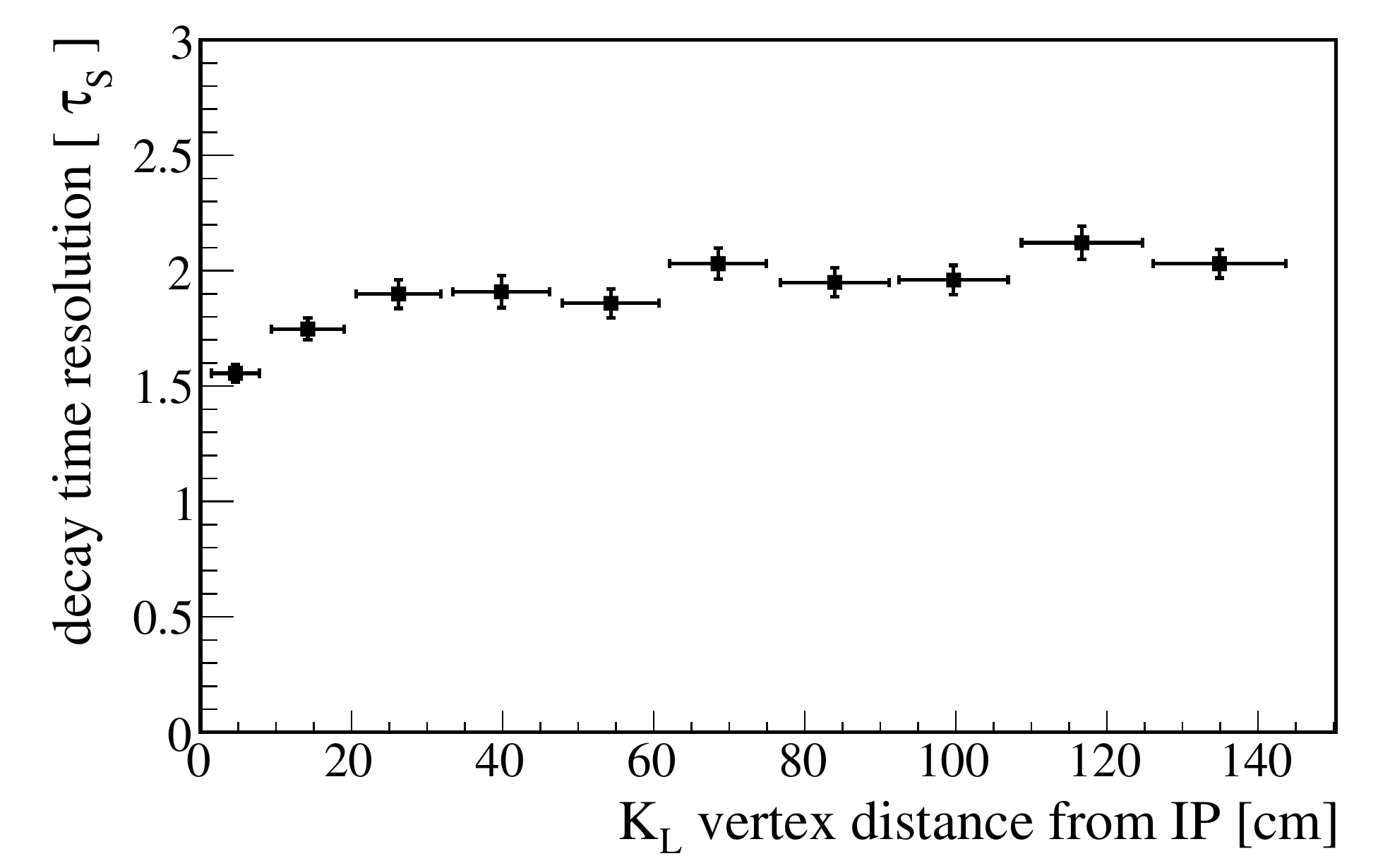}
  }
  \caption{Resolution of proper $\Kl$ decay time reconstructed for $\Kl\to3 \pi^0$ with the new method as a function of the decay vertex distance from the $\phi$ decay point (IP).}
  \label{Fig:timeres}
\end{figure}

Performance of reconstruction was tested on a sample of MC-generated $\Kl\to 3 \pi^0$ events. Resolution of proper $\Kl$ decay time was estimated for several regions of the decay vertex distance from the interaction point. Fig.~\ref{Fig:timeres} shows the resulting resolution which is at the level of $\sim 2\;\tau_S$ and remains constant with increasing $\Kl$ travelled path lengths in the whole range available in the detector. This temporal resolution is sufficient for the future \Ts~symmetry test at KLOE-2.

\section{Acknowledgements}
This work was supported in part by the EU Integrated Infrastructure Initiative Hadron Physics Project under contract number RII3-CT- 2004-506078; by the European Commission under the 7th Framework Programme through the \textit{Research Infrastructures} action of the \textit{Capacities} Programme, Call: FP7-INFRASTRUCTURES-2008-1, Grant Agreement No. 227431; by the Polish National Science Centre through the Grants No. 0469/B/H03/ 2009/37, 0309/B/H03/2011/40, 2011/03/N/ST2/02641, 2011/01/D/ST2/ 00748, 2011/03/N/ST2/02652, 2013/08/M/ST2/00323 and by the Foundation for Polish Science through the MPD programme and the project HOMING PLUS BIS/2011-4/3.
\end{document}